%% file: __main.tex
\documentclass[%
 reprint,
 superscriptaddress,
 amsmath,amssymb,
 aps,
]{revtex4-2}

\usepackage{balance}
\usepackage{graphicx}
\usepackage{bm}

\usepackage{hyperref}

\usepackage{flushend}
\usepackage{todonotes}
\usepackage{subcaption}

\usepackage{amsmath} 
\usepackage{subcaption}
\usepackage[ruled, linesnumbered]{algorithm2e}
\usepackage{braket}
\usepackage{notoccite}
\usepackage{todonotes}

\begin{document}

\title{Circuit-Level Noise Estimation via Shuttling in Plaquette Circuits}

\author{Huyen Do}
\email{huyen.do@aalto.fi}
\affiliation{Aalto University, School of Science, Department of Computer Science, Konemiehentie~2, 02150 Espoo, Finland}

\author{Alexandru Paler}
\email{alexandru.paler@aalto.fi}
\affiliation{Aalto University, School of Science, Department of Computer Science, Konemiehentie~2, 02150 Espoo, Finland}

\begin{abstract}
We present a method for estimating QEC circuit-level noise levels assuming that only single-shot measurements are available (e.g. measurements are slow and performed in a zoned/parallel fashion), and that lower level quantum hardware calibration is not possible (e.g. cloud access) or not feasible (e.g. large scale computing). We develop and run surface code plaquette experiments using two syndrome qubit configurations: FRESH, involving fresh qubits for each plaquette repetition, and RECYCLE, reusing qubits. To validate our approach, we compile plaquettes to ion-trap (IonQ Aria1) native gate set and apply hardware-aware rewrite templates to reduce circuit depth and execution time. We also run the experiments on a non-shuttling, superconducting processor (IBM Torino). We estimate circuit-level noise rates from the resulting single-shot plaquette measurement statistics, and conclude numerically about the viability of low-depth QEC experiments.
\end{abstract}

\maketitle

\section{Introduction}

Quantum algorithms are executed through quantum circuits, the backbone of quantum computing~\cite{Nielsen_Chuang_2010}. Designing efficient circuits is important, but the challenges are also related to their implementation on quantum hardware with limited fault-tolerance~\cite{eisert2025mind, preskill2025beyond}. 

In this work, building on~\cite{do2024automatic}, we study surface-code~\cite{Fowler_2012} plaquette circuits and examine how these can be used for estimating noise models. Additionally, we optimize the circuits using native-gate rewrite templates (Section~\ref{native_opt}) for ion-traps. To estimate circuit-level noise rates, we design plaquette-repetition experiments (Section~\ref{section:experiments}) using two syndrome qubit configurations: FRESH, which uses a new ancilla for each repetition, and RECYCLE, which reuses the same syndrome qubit across repetitions.

We use measurement statistics to formulate a circuit-level noise model (Section~\ref{section:error_model}). In Section~\ref{sec:res} we validate the approach on real quantum processors, including an ion-trap device (IonQ Aria1) and a superconducting device (IBM Torino).

\subsection{Circuit-level Noise and Detector Error Model}

Fault-tolerant circuits allow reliable quantum computations in the presence of noise. By introducing redundancy and measurements, errors can be detected and corrected while preserving the encoded information~\cite{Nielsen_Chuang_2010,preskill1997faulttolerant}.

\begin{figure}[!t]
    \centering
    \begin{subfigure}{0.35\linewidth}
        \centering
        \includegraphics[width=\linewidth]{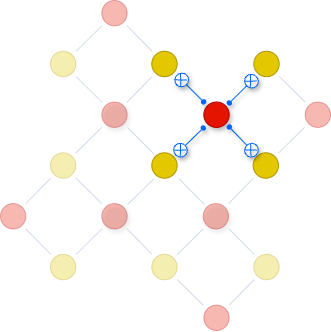}
        \caption{}
        \label{subfig:surface_code}
    \end{subfigure}
    \hfil
    \begin{subfigure}{0.35\linewidth}
        \centering
        \includegraphics[width=\linewidth]{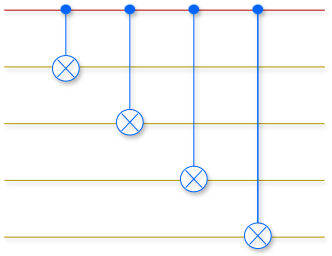} 
        \caption{}
        \label{subfig:plaquette}
    \end{subfigure}
    \caption{A plaquette in a surface code. \textbf{(a)} Mapping of a plaquette onto the lattice of the rotated distance 3 surface code, where yellow vertices are data qubits and red vertices are syndrome qubits. \textbf{(b)} Corresponding plaquette circuit.}
    \label{fig:lattice}
\end{figure}

The surface code, one of the most widely studied QEC codes, can effectively encode, detect, and correct errors if the 2-qubit quantum gate error rate is below a threshold on the order of $1\%$~\cite{Fowler_2012}. The surface code executes plaquette circuits, which measure either $\mathrm{X}$-type or $\mathrm{Z}$-type stabilizers. These two are used for detecting two primary types of errors in quantum systems: phase-flip ($\mathrm{Z}$) errors and bit-flip ($\mathrm{X}$) errors, respectively. Measuring $\mathrm{X}$($\mathrm{Z}$)-stabilizers is sensitive to $\mathrm{Z}$($\mathrm{X}$) errors because an $\mathrm{Z}$($\mathrm{X}$)-error anticommutes with a $\mathrm{X}$($\mathrm{Z}$)-stabilizer and flip its measurement outcome.

By repeatedly performing plaquette measurements over time, one obtains a stream of syndrome bit that indicate the potential presence and location of errors~\cite{dummies2023}. The exact location of the errors is not known, and the only available information are the syndrome bits. Moreover, for any syndrome bit pattern, there are multiple potential error patterns, such that it is very difficult to determine efficiently (i.e. to decode) which errors should be accounted for and corrected.

Several noise models are commonly used to characterize QEC codes, including the code-capacity, phenomenological, and circuit-level noise models~\cite{chatterjee2023qpandora,zhang_2023}.

The circuit-level noise is the most comprehensive and widely used model: errors are associated with every operation in the circuit. Noise can arise from single-qubit and two-qubit gates, initialization, reset, and readout, as well as from depolarizing errors during idle periods, crosstalk, and leakage~\cite{ktb3-gcxr}. To connect a noise model to decoding, one can translate the noisy plaquette measurement outcomes into a detector error model (DEM)~\cite{gidney2021stim,Derks_2025}. A DEM specifies a set of \emph{detectors} defined as spacetime parity checks over plaquette measurement outcomes across rounds. When a physical error occurs in the circuit, it can flip one or more stabilizer outcomes and possibly create a logical error, causing the corresponding detectors to "fire". In this way, the DEM provides a compact probabilistic description of how errors in the circuit map to detection events and logical errors. This information is used by decoders to infer and correct errors.

\subsection{Shuttling-Based QPUs}

Shuttling-based QPUs are architectures in which qubits are physically transported during computation to enable the required pairs to entangle. This capability provides reconfigurable connectivity in time. One example of shuttling-based hardware is the ion-trap quantum computer, which uses ions as qubits. Ions are trapped and manipulated using electromagnetic fields~\cite{PhysRevLett.74.4091}, and quantum information is typically encoded in their electronic or spin states. Another important example is neutral-atom devices where atoms are trapped by optical tweezers~\cite{quera,Ashkin1986SingleBeamOpticalTrap}. In such systems, qubits can be arranged across different functional regions, such as storage zones, readout zones, and entangling zones~\cite{Bluvstein2024LogicalQuantumProcessor}. 

Trout et al. (2018) investigated the practical implementation of a small-scale surface code, designated as surface-17, which consists of 9 data qubits and 8 syndrome qubits configured in a linear chain of ions~\cite{Trout_2018}. Their work is significant for illustrating how to optimize the mapping of inherently 2D plaquette circuits onto 1D ion chains, thus reducing the time needed for stabilizer measurements. Erhard et al. (2021)~\cite{erhard2021entangling} demonstrated entanglement of logical qubits through lattice surgery (LS). LS~\cite{gottesman1997stabilizer} is where groups of physical qubits arranged on the lattice can be merged and split to realize entangling gate and teleport logical information. LeBlond et al. (2023)~\cite{Leblond_2023} later introduced the Trapped-Ion Surface Code Compiler (TISCC). TISCC compiles surface code operations into circuits for trapped-ion quantum processors. The tool uses a tile-based approach to LS, focusing on local operations and optimizing for the most efficient use of resources. General compilers for shuttling based QPUs have been presented for example in~\cite{schoenberger2024shuttling, kreppel2023quantum}.

\subsection{Problem Statement}

Noise modeling is essential for evaluating and tuning the QEC performance necessary for fault-tolerant executions. In general, the distance of a QEC is determined numerically by assuming that the circuit level-noise model is already available. However, noise is fluctuating over time~\cite{proctorDetectingTrackingDrift2020, Huo_2017}, and the models might be too optimistic or pessimistic. Moreover, in the long term it will not be feasible to re-calibrate all the components of a computer for executing in real-time a fault-tolerant computation. 

It is realistic to assume a worst-case for QEC configuration purposes: we can only observe circuit execution results and have no access to previous calibration metrics, such as gate and measurement fidelities, which are anyway difficult to translate directly into the effective QEC noise parameters. Furthermore, shuttling-based architectures (e.g. ion traps and neutral atoms) introduce additional effects related to qubit movement, ancilla reuse, and waiting times that are not directly captured by standard calibration metrics. 

Our motivation is to answer the following.

\textbf{Research Question:} How can one estimate a meaningful QEC circuit-level noise model using only experimentally observable data?

In this work, we address this question through a method that uses repeated plaquette experiments using two ancilla strategies, FRESH and RECYCLE. We analyze how ancilla measurement outcomes evolve as plaquette circuits are repeated, and can estimate effective plaquette error rates and time-like correlated errors to be used in circuit-level QEC noise model estimations.

Our method does not depend heavily on measurements (i.e. is single-shot) and is based on the building blocks of the chosen QEC (e.g. surface code). Our approach is motivated by limitations of existing quantum computing platforms where dynamic circuits~\cite{corcolesExploitingDynamicQuantum2021, pinoDemonstrationTrappedionQuantum2021a} are slowing down a circuit's execution because measurement times are slow~\cite{bruzewiczTrappedionQuantumComputing2019, zhengMinimizingReadoutinducedNoise2024}, while at the same time measurements might need to be parallelized (e.g. on shuttling-based computers).

\section{Methods}

We analyze the error rates of plaquette circuits as a proxy for estimating the noise in the QPU. We introduce a protocol to estimate plaquette error rate from plaquette repetition experiments, and describe the procedures for implementing plaquette circuits on quantum processing units (QPUs). To validate our approach we chose the IonQ's Aria1 QPU~\cite{ionq_backends}. Finally, although we are motivated by shuttling-based computers, we still test the suitability of our method for non-shuttling one, such as the superconducting IBM Torino QPU~\cite{IBMQuantumTorino}.

\textbf{Definition -- plaquette error rate: } is the frequency of $1$ outcomes from the measurement of the plaquettes (i.e. measurement of syndrome qubits).

In an ideal noiseless implementation, the plaquette measurement outcomes are $0$. Therefore, a $1$ indicates that an error has occurred in the stabilizer-measurement process, which may originate from the plaquette operations: single- and two-qubit gates, ancilla idling, correlated reuse effects, or measurement noise. The plaquette error rate can be computed by increasing the number of iterations, allowing us to study how errors propagate and accumulate within the circuit. 

\textbf{Note: } Due to hardware limitations, in this work, we infer an effective error rate for a single plaquette on a specific QPU. We will extrapolate the computed error rates across multiple plaquettes when performing numerical simulations.

Plaquette-level error model is a practical metric for quantum hardware benchmarking. One can implement this model, similarly to a phenomenological one, by assuming that: a) data qubits have zero noise and b) having the measurement error rate of the syndrome qubit as a proxy for the entire plaquette. Because surface-code performance is strongly constrained by the fidelity of stabilizer measurements, comparing effective plaquette error rates across platforms directly reflects how well a device can support repeated syndrome extraction. By assessing the error rate, we can identify quantum computing platforms that deliver the highest fidelity and are capable of effectively executing surface codes.

\subsection{Plaquette Experiments}
\label{section:experiments}

We perform a comprehensive study of error accumulation within plaquette circuits on quantum hardware. Utilizing a methodology inspired by Zero Noise Extrapolation (ZNE)~\cite{Wahl_2023}, the goal is to obtain a quantitative understanding of how noise is amplified through repeated executions of plaquette circuits. Unlike ZNE, our aim is not to mitigate noise but to quantify how the reliability of a plaquette degrades as the number of repetitions within a single circuit increases.

We design an experiment of different repetitions of plaquettes for two types of syndrome qubit configurations.

\begin{figure}[t]
    \centering

    \begin{subfigure}[t]{0.38\columnwidth}
        \centering
        \includegraphics[width=\linewidth]{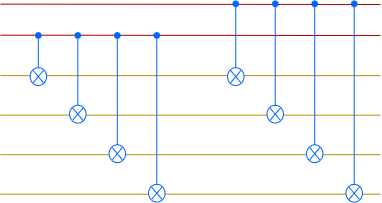}
        \caption{$n=2$}
        \label{subfig:fresh_n2}
    \end{subfigure}
    \hfill
    \begin{subfigure}[t]{0.58\columnwidth}
        \centering
        \includegraphics[width=\linewidth]{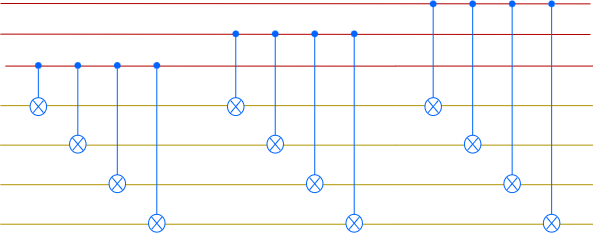}
        \caption{$n=3$}
        \label{subfig:fresh_n3}
    \end{subfigure}

    \caption{\textbf{FRESH strategy.} Each plaquette repetition uses a newly prepared syndrome qubit, so ancilla noise is largely uncorrelated across repetitions. Examples are shown for $n=2$ and $n=3$ plaquette repetitions.}
    \label{fig:typea}
\end{figure}

\textbf{Definition -- FRESH:} there are $n$ syndrome qubits used for running $n$ plaquette circuits (Fig.~\ref{fig:typea} includes definition and example).

\begin{figure}[t]
    \centering

    \begin{subfigure}[t]{0.38\columnwidth}
        \centering
        \includegraphics[width=\linewidth]{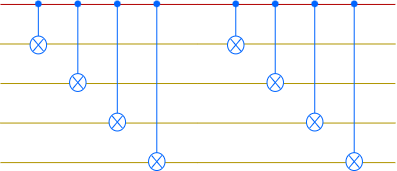}
        \caption{$n=2$}
        \label{subfig:recycle_n2}
    \end{subfigure}
    \hfill
    \begin{subfigure}[t]{0.58\columnwidth}
        \centering
        \includegraphics[width=\linewidth]{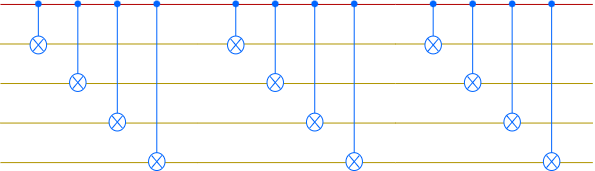}
        \caption{$n=3$}
        \label{subfig:recycle_n3}
    \end{subfigure}

    \caption{\textbf{RECYCLE strategy.} A single syndrome qubit is reused across plaquette repetitions without reset. Examples are shown for $n=2$ and $n=3$ plaquette repetitions.}
    \label{fig:typeb}
\end{figure}

\textbf{Definition -- RECYCLE:} there is a single syndrome qubit used for running $n$ plaquette circuits (Fig.~\ref{fig:typeb} includes definition and example). 

For each repetition depth $x \in \{1,\ldots, n\}$, we compile a circuit containing $x$ sequential plaquette repetitions under each strategy and execute it on hardware. From the resulting measurement histogram, we compute the probability that an ancilla measurement equals $1$ (details, including histogram processing, are in Appendix~\ref{appendix:algorithm}).

From each execution, we marginalize the histogram over data qubit outcomes and estimate the probability that a syndrome measurement equals $1$. In the FRESH configuration, a depth-$x$ circuit yields a set of probabilities $\{p_{x,y}\}_{y=1}^{x}$ where $y\in\{1,\ldots,x\}$ indexes the plaquette repetition (equivalently, the corresponding syndrome qubit). In RECYCLE, the circuit yields a single probability $p_x$ for the reused syndrome qubit.

\textbf{Note:} Our method is \emph{single-shot}. In the FRESH scenario, we perform all the measurements in parallel, in order to reduce the influence of the slow measurements, and to account for architectural constraints on shuttling QPUs. However, the syndrome qubits will be idling and introduce correlated errors.

\textbf{Note:} Because the FRESH plaquette measurements are not temporally ordered, we computed plaquette error rates instead of detection event error rates.

\begin{figure}[!t]
    \centering
    \includegraphics[width=\linewidth]{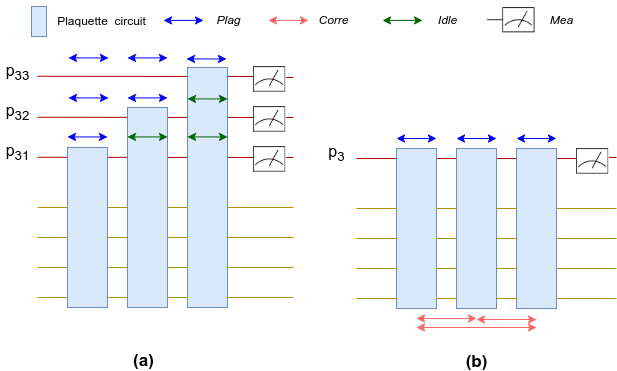}
    \caption{Experimental errors, including single plaquette errors ($plaq$), time-like correlated errors between plaquettes ($corre$), idle errors ($idle$), and measurement error ($mea$), propagating on each syndrome qubit for a circuit with $n = 3$ repetitions of the plaquette in: (a) FRESH experiment
    ; (b) RECYCLE experiment}
    \label{fig:noise_accumulate}
\end{figure}

\subsection{Developing an Error Model}
\label{section:error_model}

We develop an error model by observing the variation of the plaquette error rate.  The model parameters defined in Table~\ref{tab:symbols} are used to describe the relationships between error rates and the number of repetitions within the plaquette circuits. 

We postulate that plaquette errors are additive and that the idle error rate, a measure of error when a qubit remains idle, does not vary across multiple executions. This assumption allows us to methodically determine error rates for each plaquette circuit independently. 

In FRESH experiments, the probability $p_{xy}$ for the $y $-th ancilla reflects the cumulative error due to the sequence of $x$ plaquette circuits, the idle error that the syndrome qubit not engaging in operations and waiting for $x - y$ plaquettes to be executed, and the measurement error rate.
\begin{equation}
    p_{xy} = y \times plaq + (x-y) \times idle +  mea
\label{eq:type_a_ancilla}
\end{equation}

Conversely, in RECYCLE experiments, the probability $p_x$ must also account for the correlated errors resulting from successive operations, in addition to the plaquette and measurement errors.
\begin{equation}
    p_x = x \times plaq + corre (x-1)^{2} + mea
\label{eq:type_b_ancilla}
\end{equation}

From the relationship of Eq.\ref{eq:type_a_ancilla} and~\ref{eq:type_b_ancilla}, the correlated error $corre$ can be isolated by examining the difference between $p_x$ and $p_{xx}$, which is the probability of the $x$-th syndrome qubit being in state '1' after $x$ repetitions. By setting $x = y$, we can eliminate the idle error term.
\begin{equation}
    corre = \frac{p_x - p_{xx}}{(x-1)^2}
\label{eq:corre}
\end{equation}

\begin{table}[!t]
\centering
    \caption{Parameters of the error model.}
    \label{tab:symbols}
    \begin{tabular}{p{0.15\linewidth} p{\dimexpr0.85\linewidth-4\tabcolsep}}
    \hline
    \textbf{Symbol} & \textbf{Definition}\\
    \hline
    $n$         & Number of plaquette repetitions in the circuit. \\
    $p_x$       & Probability of '1' outcome of syndrome qubit for $x$ plaquette repetitions in RECYCLE. \\
    $p_{xy}$    & Probability of '1' outcome of the $y$-th syndrome qubit for $x$ plaquette repetitions in FRESH. \\
    $plaq$      & The error rate of a single plaquette circuit.\\
    $corre$     & The time-like correlated error rate between successive plaquettes.\\
    $idle$      & The idle error rate that an syndrome qubit waits for the execution of a plaquette circuit.\\
    $mea$       & The measurement error rate.\\
    \hline
    \end{tabular}
\end{table}

We also compute the average correlated error rate $\overline{corre}$ across all $n$ plaquette repetitions.
\begin{equation}
     \overline{corre} = \frac{1}{n-1}\sum_{i=2}^{n}\left(\frac{p_{i} - p_{ii}}{(i-1)^2}\right)
\label{eq:avg_corre}
\end{equation}

In our framework, $\overline{corre}$ is not a physical correlation coefficient; it is an effective term meant to represent a reuse-related, time-like contribution arising from pairwise correlations between any two plaquette repetitions in RECYCLE. Operationally, it is estimated from the residual difference between RECYCLE and the corresponding FRESH measurement at the same repetition depth (Eq.~\ref{eq:corre}).

Finally, we can compute $\overline{plaq}$. For FRESH, from Eq.\ref{eq:type_a_ancilla}, the average plaquette error rate is:
\begin{align}
    \overline{plaq}_{\mathrm{F}} &= \frac{1}{n-1}\sum_{i=2}^{n}(p_{ii} - p_{(i-1)(i-1)}) \\
    &= \frac{1}{n-1} (p_{nn} - p_{11})
\label{eq:plaq_FRESH}
\end{align}

For RECYCLE, the average plaquette error rate, which takes into consideration the impact of correlated errors due to ancilla reuse, is:
\begin{align}
    \overline{plaq}_{\mathrm{R}} &= \frac{1}{n-1}\sum_{i=2}^{n}(p_i - p_{i-1} - \overline{corre}) \\
    &=  \frac{1}{n-1} (p_{n} - p_{1}) -  \overline{corre}
\label{eq:plaq_RECYCLE}
\end{align}

\subsection{Circuit Optimization for Ion-Trap Native Gates}
\label{native_opt}

IonQ's quantum computers support a native gate set that includes two types of single-qubit gates (GPI, GPI2) and a two-qubit entangling gate known as the Mølmer-Sørensen (MS) gate. These gates are executed through the targeted addressing of ions using resonant lasers, which facilitate stimulated Raman transitions~\cite{IonQNativeGates}. The operational matrices for these gates are the following where $\phi_\pm = \phi_0 \pm \phi_1$:

\begin{equation}
    \small
    \setlength{\arraycolsep}{2.2pt}
    MS(\phi_0,\phi_1)
    =
    \frac{1}{\sqrt{2}}
    \begin{bmatrix}
    1 & 0 & 0 & -i e^{-i\phi_+} \\
    0 & 1 & -i e^{-i\phi_-} & 0 \\
    0 & -i e^{i\phi_-} & 1 & 0 \\
    -i e^{i\phi_+} & 0 & 0 & 1
    \end{bmatrix}
\end{equation}

\begin{equation}
    GPI(\phi) = 
    \begin{bmatrix}
        0 & e^{-i\phi} \\
        e^{i\phi} & 0 \\
    \end{bmatrix}  
\end{equation}

\begin{equation}
    GPI2(\phi) = \frac{1}{\sqrt{2}}
    \begin{bmatrix}
        1 & -i e^{-i\phi} \\
        -i e^{i\phi} & 1 \\
    \end{bmatrix}  
\end{equation}

\begin{figure}[t]
    \centering

    \begin{subfigure}[b]{0.49\columnwidth}
        \centering
        \includegraphics[width=\linewidth]{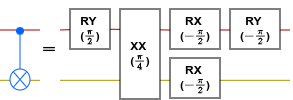}
        \caption{}
        \label{subfig:cnot_maslov}
    \end{subfigure}
    \hfill
    \begin{subfigure}[b]{0.49\columnwidth}
        \centering
        \includegraphics[width=\linewidth]{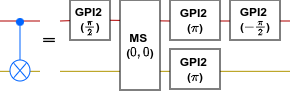}
        \caption{}
        \label{subfig:cnot_ionq}
    \end{subfigure}

    \caption{Implementation of the CNOT gate using \textbf{(a)} physical-level gates on a general ion-trap machine~\cite{Maslov_2017}, and \textbf{(b)} IonQ's native gates.}
    \label{fig:cnot_native}
\end{figure}

Given the requirements for the decomposition of a CNOT gate into ion-trap physical-level gates discussed in~\cite{Maslov_2017} (Fig.~\ref{subfig:cnot_maslov}), it is necessary to express the required rotations $ Rx\left(-\frac{\pi}{2}\right) $, $ Ry\left(-\frac{\pi}{2}\right) $, $ Ry\left(\frac{\pi}{2}\right) $ and the entangled operation $ XX\left(\frac{\pi}{4}\right) $ in terms of IonQ’s native gate set (${\mathrm{GPI}, \mathrm{GPI2}, \mathrm{MS}}$) (Fig.~\ref{subfig:cnot_ionq}). The detailed transformation is written in Appendix~\ref{appendix:native_gate_transformation}.

\begin{figure}[!t]
    \centering
    \begin{subfigure}{0.9\linewidth}
        \captionsetup{justification=centering}
        \includegraphics[width=\textwidth]{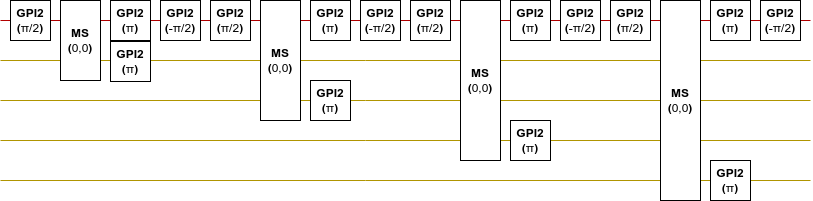}
        \caption{}
        \label{subfig:plaquette1}
    \end{subfigure}
    
    \begin{subfigure}{0.5\linewidth}
        \captionsetup{justification=centering}
        \centering
        \includegraphics[width=\textwidth]{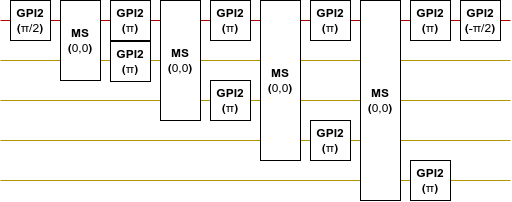}
        \caption{}
        \label{subfig:plaquette2}
    \end{subfigure}
    \hfill
    \begin{subfigure}{0.3\linewidth}
        \captionsetup{justification=centering}
        \centering
        \includegraphics[width=\textwidth]{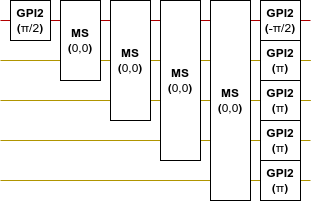}
        \caption{}
        \label{subfig:plaquette3}
    \end{subfigure}
    \caption{Evolution of the plaquette circuit optimization: \textbf{(a)} \textit{Plaquette-1}: direct transpiling from 4 CNOT gates, \textbf{(b)} \textit{Plaquette-2}: reduced circuit depth through GPI2 gate and its inverse cancellation, and \textbf{(c)} \textit{Plaquette-3}: the optimal plaquette circuit using the commutation of $\mathrm{GPI2}(\pi)$ and $\mathrm{MS}(0,0)$ gates and the cancellation of 4 consecutive $\mathrm{GPI2}(\pi)$ gates.}
    \label{fig:cnot_optimization}
\end{figure}

The experiments were conducted on a plaquette circuit consisting of four CNOT gates (Fig.~\ref{subfig:plaquette}), which is used to detect $\mathrm{Z}$ errors. We compile this plaquette directly into IonQ’s native gate set, producing the hardware-level circuit shown in Fig.~\ref{subfig:plaquette1}. We refer to this unoptimized native implementation as \textit{Plaquette-1}.

\begin{figure}[t]
    \centering
    
    \begin{subfigure}[t]{0.48\columnwidth}
        \centering
        \includegraphics[width=\linewidth]{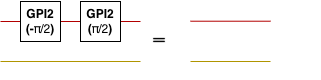}
        \caption{}
        \label{subfig:cancel_pair}
    \end{subfigure}
    \hfill
    \begin{subfigure}[t]{0.48\columnwidth}
        \centering
        \includegraphics[width=\linewidth]{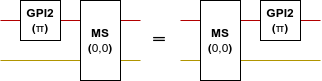}
        \caption{}
        \label{subfig:commute_ms}
    \end{subfigure}

    \vspace{0.6em}

    \begin{subfigure}[t]{0.75\columnwidth}
        \centering
        \includegraphics[width=\linewidth]{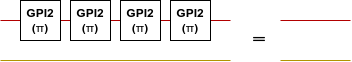}
        \caption{}
        \label{subfig:cancel_four}
    \end{subfigure}

    \caption{Circuit rewrite templates.
    \textbf{(a)} $\mathrm{GPI2}(-\pi/2)$ followed by $\mathrm{GPI2}(\pi/2)$ cancels to identity.
    \textbf{(b)} $\mathrm{GPI2}(\pi)$ on the first qubit commutes with $\mathrm{MS}(0,0)$.
    \textbf{(c)} Four consecutive $\mathrm{GPI2}(\pi)$ gates cancel to identity.}
    \label{fig:rewrite_rule}
\end{figure}

To reduce circuit depth, we developed a template rewrite rule based on the characteristic of the GPI2 gate, which states $ GPI2^{-1}(\phi) = GPI2(\phi + \pi) $. This rule enables the elimination of consecutive GPI2 gates that are inverse to each other (Fig.~\ref{subfig:cancel_pair}). Employing this optimization yields a shorter version of the original circuit (\textit{Plaquette-2} from Fig.~\ref{subfig:plaquette2}). Upon analyzing \textit{Plaquette-2}, a commutation is observed between the MS(0,0) and GPI2($\pi$) gates (Fig.~\ref{subfig:commute_ms}). By applying this rule, all single-qubit GPI2($\pi$) gates are shifted to the right, leading to the discovery that a sequence of four GPI2($\pi$) gates is equivalent to the identity operation (Fig.~\ref{subfig:cancel_four}). Consequently, the optimal version of the plaquette circuit, called \textit{Plaquette-3} (Fig.~\ref{subfig:plaquette3}), is derived. The proofs for the rewrite templates are written in the Appendix~\ref{appendix:proof}.

\section{Results}
\label{sec:res}

We analyze the plauquette error rate in the FRESH and RECYCLE experiments, conducted on IonQ Aria1 (trapped-ion) and IBM Torino (superconducting). Using the error-accumulation model developed in Section~\ref{section:error_model}, we convert the collected measurement probabilities into estimates of the plaquette error rate under two strategies ($\overline{plaq}_{\mathrm{F}}$, $\overline{plaq}_{\mathrm{R}}$), and time-like correlated error ($\overline{corre})$.

The results in Table~\ref{table:error_rate} highlight the impact of circuit optimization and also indicate that the correlation term behaves differently across architectures, reflecting differences between reconfigurable-connectivity (shuttling-style) platforms and fixed-layout devices. All plots from the experiments are provided in Appendix~\ref{appendix:plots}.

\begin{figure}[!t]
    \centering
    \includegraphics[width=0.99\linewidth]{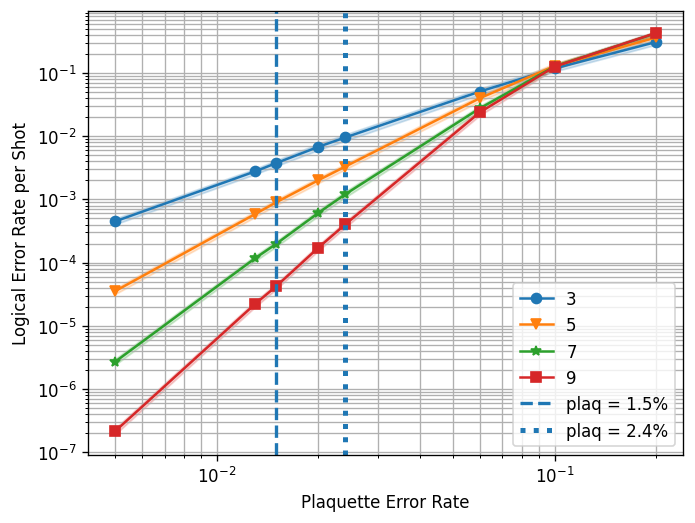}
    \caption{Logical error rate of rotated surface code (distances 3, 5, 7, 9) using an effective plaquette error rate $\overline{plaq}=1.5\%$ and a naive gate-sum baseline of $10\%$.}
    \label{fig:stim_simulation}
\end{figure}

We obtained a) Plaquette-1 -- 2.5\%, b) Plaquette-2 -- 2.4\%, c) Plaquette-3 -- 1.3\%. In Fig.~\ref{fig:stim_simulation} we plotted two horizontal lines to indicate how the ion-trap QPU would perform in experiments at the predicted error rates.

\textbf{Note:} Fig.~\ref{fig:stim_simulation} should be used with caution, because we simulate the execution of multiple plaquettes based on the results of running a single plaquette. Moreover, due to the prohibitive cost of running long experiments, our $\overline{plaq}$ were obtained using a few data points.

\begin{table}[!t]
    \centering
    \captionsetup{justification=centering}
    \begin{tabular}{l c c c c c c}
    \hline
    \textbf{Setting} 
    & \multicolumn{3}{c}{\textbf{IonQ Aria1}} 
    & \multicolumn{3}{c}{\textbf{IBM Torino}} \\
    \textbf{} 
    & $\overline{plaq}_{\mathrm{F}}$ & $\overline{plaq}_{\mathrm{R}}$ & $\overline{corre}$
    & $\overline{plaq}_{\mathrm{F}}$ & $\overline{plaq}_{\mathrm{R}}$ & $\overline{corre}$ \\
    \hline
    Plaq.-1/Opt-0 & 3.0 & 2.9 & 0.9 & 3.2 & 2.9 & -1.5 \\
    Plaq.-2/Opt-1 & 2.3 & 2.5 & 0.5 & 1.7 & 0.6 & -0.6 \\
    Plaq.-3/Opt-2 & 1.5 & 1.6 & -0.1   & 1.5 & 0.9 & -0.8 \\
    \hline
    \end{tabular}

    \caption{Estimated plaquette error rates (\%) on IonQ Aria1 and IBM Torino. For IonQ, rows correspond to three plaquette circuit variants. For IBM, rows correspond to circuits with different compilation optimization levels.}
    \label{table:error_rate}
    
\end{table}

\subsection{IonQ Aria1 QPU}

From Table~\ref{table:error_rate}, we observe a consistent reduction in the estimated plaquette failure rate across the three circuit variants, with \textit{Plaquette-3} exhibiting the lowest error, followed by \textit{Plaquette-2} and \textit{Plaquette-1}. This trend highlights the benefit of native-gate-level circuit optimization. For the most optimized implementation, we obtain $\overline{plaq}_{\mathrm{F}} = \overline{plaq}_{\mathrm{R}} \approx 1.5\%$.

At the time the experiments were run (April 2024), the randomized benchmarking calibration metrics for Aria1 were 2.5\% for two-qubit gate error and 0.03\% for single-qubit gate error. A naive upper bound based on summing two-qubit gate error rates (four entangling gates at $\approx 2.5\%$ each) would suggest a much larger error than observed here. However, these quantities are not directly comparable: randomized benchmarking reports an average per-gate error, while $\overline{plaq}$ is an effective circuit-level noise error rate. In particular, syndrome outcome $1$ is only a partial signature of errors in the stabilizer-measurement process; circuit structure and error propagation can lead to an effective plaquette error that differs significantly from a linear sum of per-gate error rates.

\subsection{IBM Torino QPU}

Table~\ref{table:error_rate} shows negative values of $\overline{corre}$ for IBM Torino across the optimization levels. A negative $\overline{corre}$ indicates that, for the same number of plaquette repetitions, the probability that the syndrome measurement outcome is $1$ grows more slowly in RECYCLE than in FRESH.

\begin{figure}[!t]
    \begin{subfigure}{0.35\linewidth}
        \includegraphics[width=\linewidth]{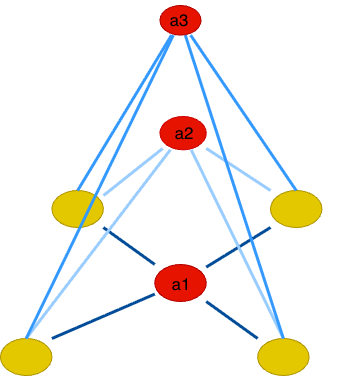}
        \caption{}
        \label{subfig:shuttling_fresh}
    \end{subfigure}
    \begin{subfigure}{0.6\linewidth}
        \includegraphics[width=\linewidth]{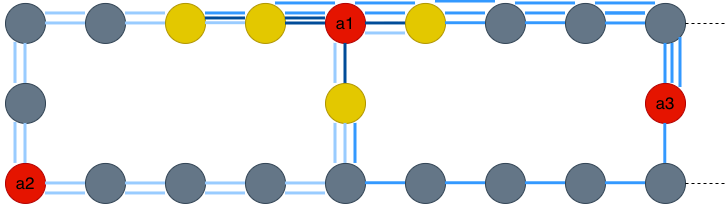}
        \caption{}
        \label{subfig:fixed_fresh}
    \end{subfigure}
    \caption{Illustration of FRESH strategy under two connectivity models. (a) Shuttling-based QPU: fresh qubits $a_1, a_2, a_3$ can be transported and directly entangled with data qubits. (b) Fixed-layout-based QPUs: ancillas are chosen from a static coupling graph of qubits in the device, so introducing fresh qubits requires additional routing two-qubit gates (blue lines) to realize the same interaction pattern.}
    
\end{figure}  

A plausible explanation is architectural: IBM Torino has a fixed 2D connectivity graph, so implementing the FRESH strategy would require extra routing operations, often SWAP gate insertions, to perform the required interactions. These additional two-qubit operations increase circuit depth and noise exposure, increasing the measured syndrome probability in FRESH. In contrast, RECYCLE reuses a single syndrome and therefore admits a more local mapping with lower routing overhead, leading to smaller observed error accumulation. Under these conditions, the RECYCLE--FRESH difference used to estimate $\overline{corre}$ no longer isolates a reuse-induced time-like correlated effect between plaquettes. Consequently, negative $\overline{corre}$ suggests that our circuit-level plaquette noise model does not apply to fixed-layout QPUs.

\section{Discussion}

The results in Table~\ref{table:error_rate} highlight two main observations. First, native-gate-level circuit rewrite-rule optimization improves plaquette reliability. On IonQ Aria1, the estimated plaquette failure rate decreases from \textit{Plaquette-1} to \textit{Plaquette-3}, with an overall reduction of the order of a factor of two. On IBM Torino, we observe a similar decrease in the estimated plaquette failure rate as the transpiler optimization level increases. This trend appears consistently under both the FRESH and RECYCLE strategies.

\subsection{Temporal Evolution of Noise on Ion-Trap QPUs}

The average plaquette error rate $\overline{plaq}$ provides a practically useful abstraction: it compresses device- and circuit-dependent behavior into a single parameter that can be directly incorporated into simulations of quantum error correction codes. In Fig.~\ref{fig:stim_simulation}, we use $\overline{plaq}$ as an effective plaquette-measurement noise parameter in surface QEC simulations on Aria1. 

For Aria1 we can notice that the measurement error rates drop significantly after approximately three executions of the plaquette circuit. For example, repeating FRESH for four times will require four syndrome qubits. Each ancilla will be entangled, one after the other, and then it will idle until all of the syndromes are measured (e.g. the first ancilla will idle for the time it takes to run the next three plaquettes, the second ancilla will idle for the time it takes to run the following two plaquettes -- see Fig.~\ref{fig:noise_accumulate} and Fig.~\ref{fig:ionq_refresh}).

As a result, while idling, the ancillas will also collect noise from the data qubits. It is not only the idling errors that are affecting the measurement statistics of the ancillas, but also the noise accumulating on the data qubits. Consequently, the ancillas used in the FRESH scheme are a proxy for the exponential noise on the data qubits. Therefore, we consider the drop of the measurement error rate as an indication that both the ancilla and the data qubits have decohered beyond recovery. To this end, for computing the average plaquette error rate, we considered only three time steps. For $n$ time steps, we have (n-2) 3-step windows: ((1,2,3), (2,3,4), \ldots, (n-2,n-1,n)). The corresponding early-time FRESH estimator from Eq.~\ref{eq:plaq_FRESH} is now: 
\begin{align}
    \overline{plaq}_{\mathrm{F}} &= \frac{1}{n-2}\sum_{i=1}^{n-2} \frac{(p_{(i+2)(i+2)} - p_{ii})}{2}
\label{eq:plaq_FRESH}
\end{align}

\subsection{Correlated Errors on Superconducting QPUs}

The behavior of the time-like correlated term $\overline{corre}$ clarifies the architectural regime in which this circuit-level noise model is the most meaningful. The model is designed to separate per-plaquette error contributions from time-like accumulation effects introduced when the same physical ancilla is reused across repetitions. These effects can appear as pairwise correlations between plaquette repetitions. This separation is most natural on shuttling-style platforms, where qubit reuse and waiting are explicit scheduling primitives: an ancilla can be repeatedly brought into and out of interaction regions (Fig.~\ref{subfig:shuttling_fresh}). In contrast, on fixed-layout superconducting devices, implementing the FRESH strategy typically introduces additional routing overhead (Fig.~\ref{subfig:fixed_fresh}). As a result, the RECYCLE--FRESH difference no longer isolates reuse-induced time-like effects, which is consistent with the negative $\overline{corre}$ values observed on IBM Torino. 

\section{Conclusion}

We presented a method for estimating circuit-level noise from experiments that repeat QEC plaquette circuits on quantum hardware. By comparing two syndrome qubit configurations, FRESH, which uses a fresh ancilla for each repetition, and RECYCLE, which reuses the same qubit, we infer an effective plaquette error rate that can be directly incorporated into surface-code simulations. This provides a circuit-level alternative to constructing noise models solely from gate-level calibration metrics.

Our approach offers a practical tool for benchmarking QEC primitives on near-term quantum hardware. On the shuttling-based IonQ Aria1 trapped-ion processor, hardware-aware native-gate rewrite templates reduced circuit depth and consistently improved plaquette reliability across optimized circuit variants. On the superconducting IBM Torino processor, which has a fixed two-dimensional qubit layout, the inferred inter-plaquette correlation term suggests that routing and compilation overhead can dominate the RECYCLE–FRESH difference.

Our noise-estimation method based on plaquettes is compatible with shuttling-style platforms, where qubit reuse and waiting are explicit scheduling primitives that can induce time-like correlations across repeated plaquette execution. Our method provides a hardware-aware way to connect experimentally measured QEC primitives with circuit-level surface-code simulations.

The source code used in this work is \url{https://github.com/huyenemma/qt-native-gate-optimization}. In addition, our IonQ native-gate rewrite rules are implemented at \url{https://github.com/ionq-samples/Ion-Q-Thruster}. 

\begin{acknowledgments}
The authors gratefully acknowledge IonQ, Inc. and IBM Quantum for providing access to their quantum hardware.
\end{acknowledgments}

\bibliography{__main}

\appendix
\input{__appendix}

\end{document}

%% file: __appendix.tex
\section{Superconducting QPUs}

We use superconducting QPUs as a comparison platform to evaluate the generality of our noise-estimation method. Superconducting quantum computers utilize circuits cooled to near absolute zero to create controllable quantum states~\cite{Kockum2019, wu2021mapping}.

In contrast to ion-trap quantum computers, superconducting processors typically have a fixed, planar nearest-neighbor connectivity graph, meaning that only adjacent qubits can be entangled directly. Recent superconducting chips are 2.5D and include also longer range connections. Therefore, when a logical two-qubit gate involves non-adjacent qubits, additional routing operations (e.g $\mathrm{SWAP}$ gates) must be inserted to bring them next to each other. These extra two-qubit gates increase circuit depth, which in turn amplifies the impact of physical error rates. Successful execution on superconducting hardware, therefore, requires compilation strategies that take both the device connectivity and native gate set into account, with the goal of minimizing routing overhead and reducing accumulated error.

\section{Circuit Equivalence through Template Rewrite Rules}
\label{appendix:eq}

\begin{figure}[!ht]
    \centering
    \begin{subfigure}[b]{0.4\textwidth}
        \centering
        \includegraphics[scale=0.45]{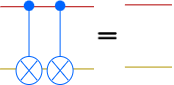}
        \caption{}
    \end{subfigure}
    \begin{subfigure}[b]{0.4\textwidth}
        \centering
        \includegraphics[scale=0.45]{img/opt2.drawio.png}
        \caption{}
    \end{subfigure}
    \caption{Example circuit rewrite templates. 
    \textbf{(a)} Cancellation of two consecutive CNOT gates. 
    \textbf{(b)} Commutation of \(\mathrm{GPI2}(\pi)\otimes I\) with \(\mathrm{MS}(0,0)\).}
    \label{fig:rewrite_examples}
    \label{fig:cnot_combined}
\end{figure}

Template rewrite rules are a circuit-optimization technique based on circuit equivalence: different gate sequences can implement the same operation and are therefore interchangeable. A \textit{template} matches a sequence of gates and substitutes them with an alternative sequence~\cite{1219016} that does the same transformation but is cheaper under some cost metric, such as fewer gates, smaller depth, or shorter execution time. The rules are implemented by their basis on algebraic identities~\cite{Hietala_2021} that define how certain combinations of quantum gates can be replaced by simpler equivalents or even eliminated. For instance, two consecutive CNOT (CX) gates cancel each other out (Fig.~\ref{fig:cnot_combined}a).

In practice, template rewriting scans a circuit for occurrences of \textit{templates}. Because rewrites are local, they are typically applied iteratively: each subsitution introduce a new circuit that may expose further simplifications. Depending on the optimization objective, the same equivalence can be used in different ways, for example, to reduce gate counts or restructure the circuit to better match a target hardware.

\section{Rotation around basic axis gates}
\label{appendix:rotation_gate}

\noindent \textbf{RX}: Setting $\phi = 0 $ obtains the rotation by the angle $\theta$ around X axis:  
\begin{equation}
    RX(\theta) = 
\begin{pmatrix}
    \cos\frac{\theta}{2} & -i\sin\frac{\theta}{2} \\
    -i\sin\frac{\theta}{2} & \cos\frac{\theta}{2}
\end{pmatrix}
= R(\theta, 0).
\end{equation}

\noindent \textbf{RY}: Setting $\phi = \frac{\pi}{2} $ gets the rotation by the angle $\theta$ around Y axis:

\begin{equation}
    RY(\theta) = 
\begin{pmatrix}
    \cos\frac{\theta}{2} & -\sin\frac{\theta}{2} \\
    \sin\frac{\theta}{2} & \cos\frac{\theta}{2}
\end{pmatrix}
= R\left(\theta, \frac{\pi}{2}\right).
\end{equation}

\noindent \textbf{RZ}: Cannot obtain the rotation about Z axis by a single rotation R pulse, therefore require a circuit of two or more rotation gate. 

\begin{equation} 
    RZ(\theta) = RY\left(\frac{-\pi}{2}\right) \cdot RX(\theta) \cdot RY\left(\frac{\pi}{2}\right) =
\begin{pmatrix}
    e^{-i\theta/2} & 0 \\
    0 & e^{i\theta/2}
\end{pmatrix}
\end{equation}

\section{Transformation to IonQ's native gates}
\label{appendix:native_gate_transformation}
\begin{equation}
    XX\left(\frac{\pi}{4}\right) = \frac{1}{\sqrt{2}}
    \begin{bmatrix}
        1 & 0 & 0 & -i \\
        0 & 1 & -i & 0 \\
        0 & -i & 1 & 0 \\
        -i & 0 & 0 & 1 \\
    \end{bmatrix}
    = MS(0, 0)
\end{equation}

\begin{equation}
    RX\left(-\frac{\pi}{2}\right) = \frac{1}{\sqrt{2}}
    \begin{bmatrix}
        1 & i \\
        i & 1 \\
    \end{bmatrix}
    = GPI2(\pi)
\end{equation}

\begin{equation}
    RY\left(-\frac{\pi}{2}\right) = \frac{1}{\sqrt{2}}
    \begin{bmatrix}
        1 & 1 \\
        -1 & 1 \\
    \end{bmatrix}
    = GPI2\left(\frac{\pi}{2}\right)
\end{equation}

\begin{equation}
    RY\left(\frac{\pi}{2}\right) = \frac{1}{\sqrt{2}}
    \begin{bmatrix}
        1 & -1 \\
        1 & 1 \\
    \end{bmatrix} 
    = GPI2\left(-\frac{\pi}{2}\right)
\end{equation}

\section{Mathematical proofs of the native gate rewrite templates}
\label{appendix:proof}

\subsection{Cancellation of two consecutive GPI2 gates}

Using the definition of $\mathrm{GPI2}$, we have
\begin{equation}
    \begin{aligned}
        \mathrm{GPI2}\left(-\frac{\pi}{2}\right)\mathrm{GPI2}\left(\frac{\pi}{2}\right)
        &=
        \left(
        \frac{1}{\sqrt{2}}
    \begin{bmatrix}
        1 & i \\
        i & 1
    \end{bmatrix}
    \right)
    \left(
    \frac{1}{\sqrt{2}}
    \begin{bmatrix}
        1 & -i \\
        -i & 1
    \end{bmatrix}
    \right) \\[0.5em]
    &=
    \frac{1}{2}
    \begin{bmatrix}
        1+1 & -i+i \\
        i-i & 1+1
    \end{bmatrix} \\[0.5em]
    &=
    \begin{bmatrix}
        1 & 0 \\
        0 & 1
    \end{bmatrix}
    =
    \mathbb{I}.
    \end{aligned}
\end{equation}

\subsection{Commutation of \(\mathrm{GPI2}(\pi)\otimes I\) with \(\mathrm{MS}(0,0)\)}

Let
\[
A = \mathrm{GPI2}(\pi)\otimes I
=
\frac{1}{\sqrt{2}}
\begin{bmatrix}
    1 & 0 & i & 0 \\
    0 & 1 & 0 & i \\
    i & 0 & 1 & 0 \\
    0 & i & 0 & 1
\end{bmatrix}
\]
and
\[
M = \mathrm{MS}(0,0)
=
\frac{1}{\sqrt{2}}
\begin{bmatrix}
    1 & 0 & 0 & -i \\
    0 & 1 & -i & 0 \\
    0 & -i & 1 & 0 \\
    -i & 0 & 0 & 1
\end{bmatrix}.
\]
Then
\begin{equation}
\begin{aligned}
AM
&=
\frac{1}{2}
\begin{bmatrix}
    1 & 1 & i & -i \\
    1 & 1 & -i & i \\
    i & -i & 1 & 1 \\
    -i & i & 1 & 1
\end{bmatrix}, \\[0.5em]
MA
&=
\frac{1}{2}
\begin{bmatrix}
    1 & 1 & i & -i \\
    1 & 1 & -i & i \\
    i & -i & 1 & 1 \\
    -i & i & 1 & 1
\end{bmatrix}.
\end{aligned}
\end{equation}
Therefore,
\[
(\mathrm{GPI2}(\pi)\otimes I)\mathrm{MS}(0,0)
=
\mathrm{MS}(0,0)(\mathrm{GPI2}(\pi)\otimes I).
\]

\subsection{Four consecutive \(\mathrm{GPI2}(\pi)\) gates cancel up to a global phase.}

We first compute
\begin{equation}
    \begin{aligned}
    \mathrm{GPI2}(\pi)^2
    &=
    \frac{1}{2}
    \begin{bmatrix}
        1 & i \\
        i & 1
    \end{bmatrix}
    \begin{bmatrix}
    1 & i \\
    i & 1
    \end{bmatrix} \\[0.4em]
    &=
    \frac{1}{2}
    \begin{bmatrix}
    0 & 2i \\
    2i & 0
    \end{bmatrix}
    =
    \begin{bmatrix}
    0 & i \\
    i & 0
    \end{bmatrix}.
    \end{aligned}
\end{equation}
Therefore,
\begin{equation}
\begin{aligned}
\mathrm{GPI2}(\pi)^4
&=
\left(\mathrm{GPI2}(\pi)^2\right)^2 \\[0.4em]
&=
\begin{bmatrix}
    0 & i \\
    i & 0
\end{bmatrix}
\begin{bmatrix}
    0 & i \\
    i & 0
\end{bmatrix} \\[0.4em]
&=
\begin{bmatrix}
    -1 & 0 \\
    0 & -1
\end{bmatrix}
=
-\mathbb{I}.
\end{aligned}
\end{equation}

\section{Details of extracting \(p_x\), \(p_{xy}\) from histogram}
\label{appendix:algorithm}

\begin{algorithm}[!t]
\caption{Algorithm to estimate the plaquette failure rate.}
\label{algo:error}

$N \gets$ number of repetitions\;
\For{type $\in \{ \text{FRESH}, \text{RECYCLE} \}$}{
   \For{$depth \gets 1$ \KwTo $N$}{
        Compile circuit consisting of plaquette $\times$ depth based on type\;
        Execute circuit\;
        Collect the measurement histogram of bitstrings\;
        Bitstrings $\gg$ 4\;
        \uIf{type $=$ FRESH}{
            \For{$i \gets 0$ \KwTo $depth$}{
                Bitstrings $\ll i \gg$ ($depth - 1 - i$)\;
                Collect probability of state $\ket{1}$\;
            }
        }\ElseIf{type $=$ RECYCLE}{
            Compute probability of state $\ket{1}$\;
        }
   }
}
\end{algorithm}

\begin{figure*}[!t]
    \centering
    \begin{subfigure}[b]{0.32\textwidth}
        \centering
        \captionsetup{justification=centering}
        \includegraphics[width=\textwidth]{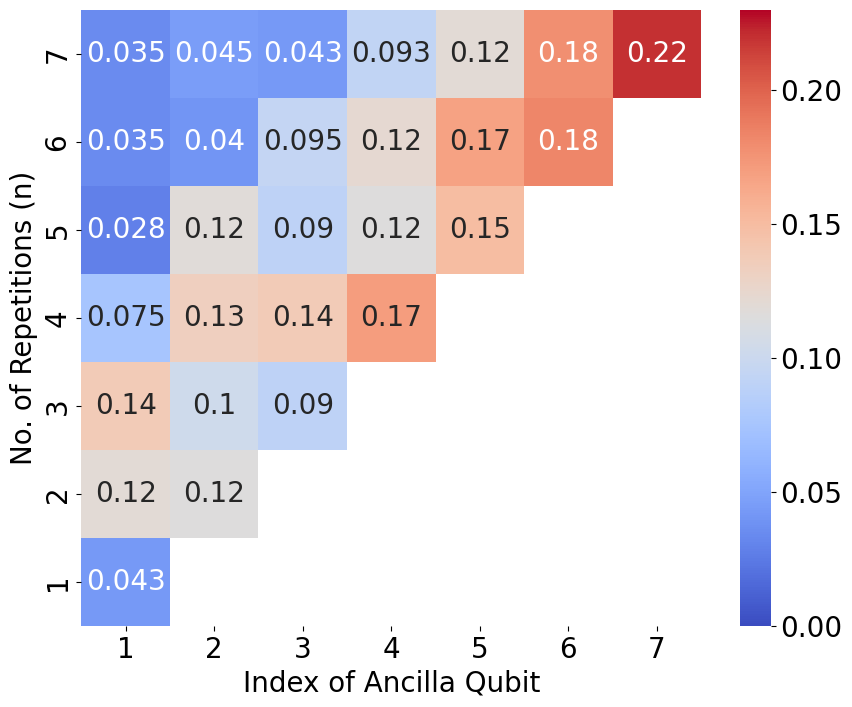}
        \caption{}
        \label{fig:a12}
    \end{subfigure}
    \begin{subfigure}[b]{0.32\textwidth}
        \centering
        \captionsetup{justification=centering}
        \includegraphics[width=\textwidth]{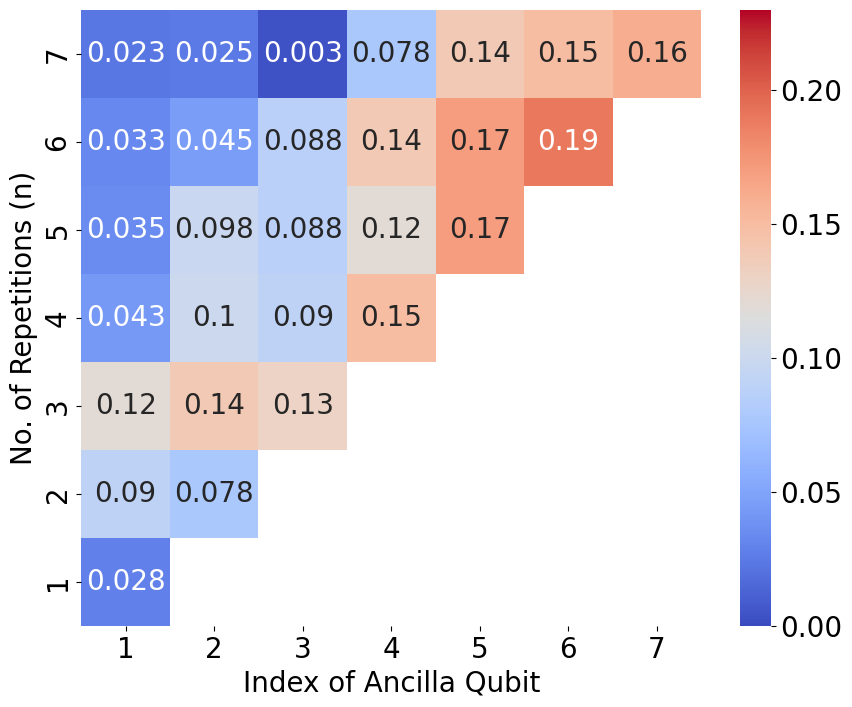}
        \caption{}
        \label{fig:a22}
    \end{subfigure}
    \begin{subfigure}[b]{0.32\textwidth}
        \centering
        \captionsetup{justification=centering}
        \includegraphics[width=\textwidth]{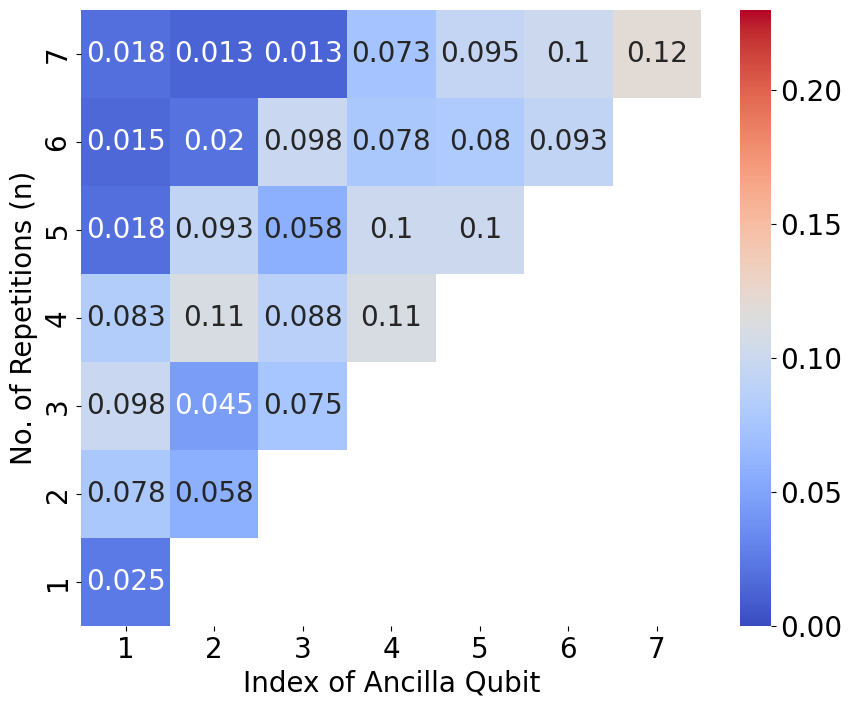}
        \caption{}
        \label{fig:a32}
    \end{subfigure}
    \caption{FRESH experiment on IonQ Aria1: the probability that the ancilla measurement outcome is $1$ versus circuit depth for \textbf{(a)} Plaquette-1, \textbf{(b)} Plaquette-2, and \textbf{(c)} Plaquette-3} 
    \label{fig:ionq_refresh}
\end{figure*}

Algorithm~\ref{algo:error} is designed to estimate the failure rate of plaquette circuits. 

\begin{enumerate}
    \item \textbf{Line 4:} Depending on the experiment type (FRESH or RECYCLE), the algorithm generates circuits with a specific number of repetitions of plaquettes. This step ensures that the circuits are tailored to the specific experimental conditions.
    
    \item \textbf{Line 5:} The generated circuits are then sent to the IonQ hardware API for execution. Measurement gates are placed at the end of the circuit for every qubit. 

    \item \textbf{Line 6:} After executing the circuits, the algorithm collects the measurement histogram, which provides a statistical representation of the frequency of different measurement outcomes. Each measurement outcome corresponds to a bitstring representing the states of the qubits in the circuit.  The histogram \{bitstring: frequency\} records the occurrence frequency of each bitstring observed during measurement.
    
    \item \textbf{Line 7:} Since the focus is on analyzing the measurement outcomes of the ancilla qubits only, the algorithm shifts the bitstrings four digits to the right. This operation effectively eliminates the measurement outcomes associated with the four data qubits in the circuit, isolating the measurement results of the ancilla qubits. For example, for a measurement outcome of a single plaquette (1 ancilla and 4 data qubits) represented by the bitstring "11011", shifting the bitstring four digits to the right results in "1", indicating the measurement outcome of the ancilla qubit.
    
    \item \textbf{Line 10:} If executing a FRESH experiment with multiple ancilla qubits, the algorithm aims to extract the probability of each ancilla qubit yielding the measurement outcome "1". This is achieved by effectively shifting the digits of the bitstring to isolate the outcome of each ancilla qubit. For instance, in a depth-2 circuit with 2 ancilla qubits, after line 7, the outcome is "10". The outcome of the first ancilla is obtained by shifting the bitstring by \(0\) digits: "10" \(<<\) \(0\) \(>>\) \((2 - 1 - 0)\) = "10" \(>>\) \(1\) = "1". Similarly, the outcome of the second ancilla is obtained by shifting the bitstring by \(1\) digit: "10" \(<<\) \(1\) \(>>\) \((2 - 1 - 1)\) = "0" \(>>\) \(0\) = "0".
    
   \item \textbf{Line 11:} After isolating the measurement outcomes of the ancilla qubits, the algorithm computes the probability of obtaining the measurement outcome "1" for each ancilla qubit. This is achieved by summing up the frequencies of the histogram corresponding to the measurement outcome "1".

    \item \textbf{Line 14:} If executing a RECYCLE experiment, where each circuit involves only one qubit, no further shifting is needed as there is only one measurement outcome to consider.
\end{enumerate}

\section{Experiment Results}
\label{appendix:plots}

\subsection{IonQ Aria1 QPU}

For IonQ Aria1, we report the probability that the ancilla measurement outcome is $1$ as a function of circuit depth (number of plaquette repetitions). We evaluate three plaquette circuit variants (Plaquette-1 to Plaquette-3) under both ancilla strategies, FRESH and RECYCLE. All circuits were run with 1000 shots.

\textbf{FRESH Experiment.} Figure~\ref{fig:ionq_refresh} presents the results conducted across three different plaquettes on the Aria1 QPU. Each part of the figure corresponds to outcomes from Plaquettes 1, 2, and 3, revealing the probability of error for each ancilla qubit. On the heatmaps, the horizontal axis specifies the ancilla qubits, while the vertical axis counts the repetitions of the plaquettes within a circuit. The visual transition from cooler blues to warmer reds within these heatmaps signifies an increasing likelihood of errors. Notably, \textit{Plaquette-3} exhibits significantly fewer errors than the other two plaquettes, indicating the effectiveness of optimization applied to the plaquette design.

\begin{figure*}[!t]
    \begin{subfigure}[b]{0.32\textwidth}
        \centering
        \captionsetup{justification=centering}
        \includegraphics[width=\textwidth]{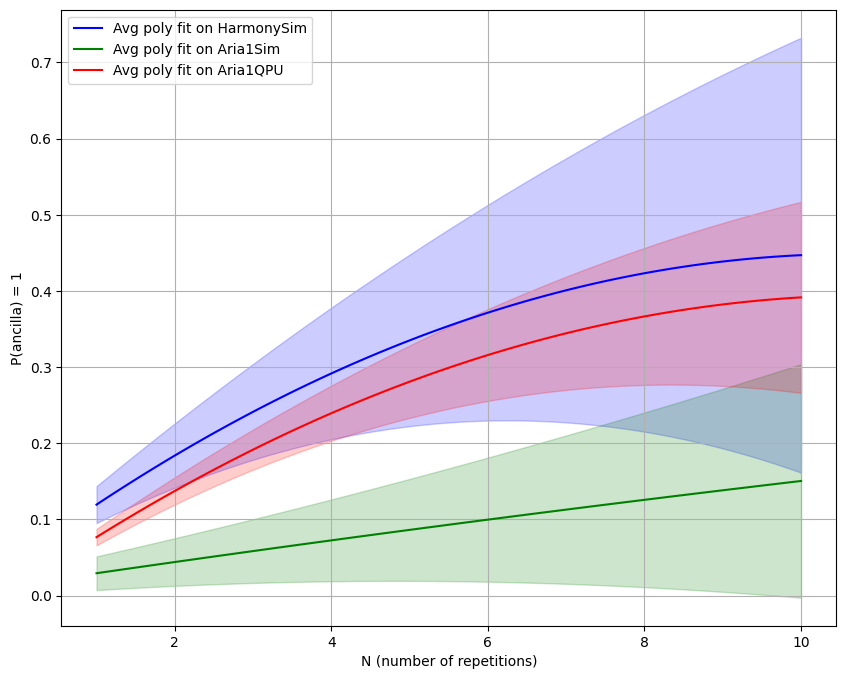}
        \caption{}
        \label{fig:b1}
    \end{subfigure}
    \begin{subfigure}[b]{0.32\textwidth}
        \centering
        \captionsetup{justification=centering}
        \includegraphics[width=\textwidth]{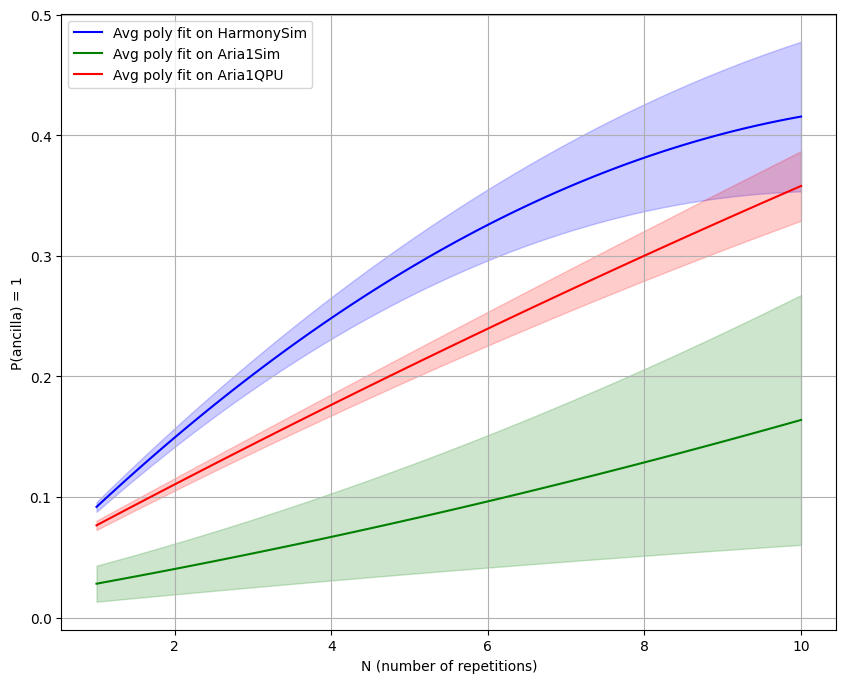}
        \caption{}
        \label{fig:b2}
    \end{subfigure}
    \begin{subfigure}[b]{0.32\textwidth}
        \centering
        \captionsetup{justification=centering}
        \includegraphics[width=\textwidth]{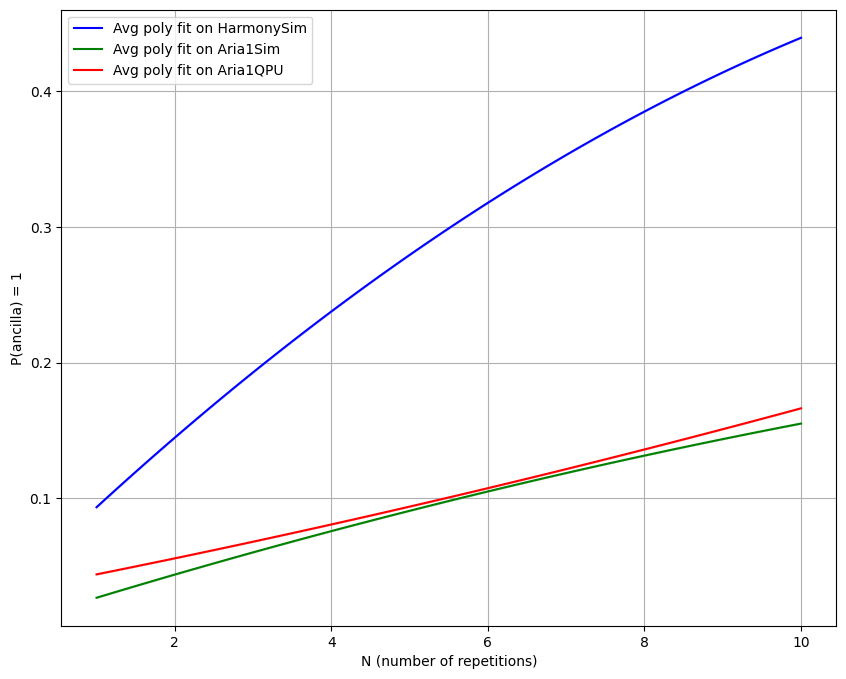}
        \caption{}
        \label{fig:b3}
    \end{subfigure}
    \caption{RECYCLE experiment on IonQ Aria1: polynomial fits (with confidence intervals) of the probability that the ancilla measurement outcome is $1$ versus circuit depth for \textbf{(a)} Plaquette-1, \textbf{(b)} Plaquette-2, and \textbf{(c)} Plaquette-3. Confidence intervals are derived from three independent trial runs.}
    \label{fig:ionq_recycle}
\end{figure*}

\textbf{RECYCLE Experiment.} Figure~\ref{fig:ionq_recycle} illustrates degree 2 fit curves with confidence intervals obtained from three separate runs, each consisting of 1000 shots, to display the probability of RECYCLE ancilla qubit registering '1' for \textit{Plaquette-1}, 2, and 3.  These probabilities are plotted as functions of the number of plaquette repetitions within a circuit on both noisy simulators and the actual QPU. Notably, the blue line represents results from the Harmony Simulator (with 11 algorithmic qubits), the green line from the Aria1 Simulator (with 25 algorithmic qubits), and the red line from the QPU Aria1, presumed to have a noise profile broadly similar to its simulator. 

\subsection{IBM Torino QPU}

\begin{figure*}[!h]
    \centering
    \begin{subfigure}[t]{0.31\textwidth}
        \centering
        \includegraphics[width=\linewidth]{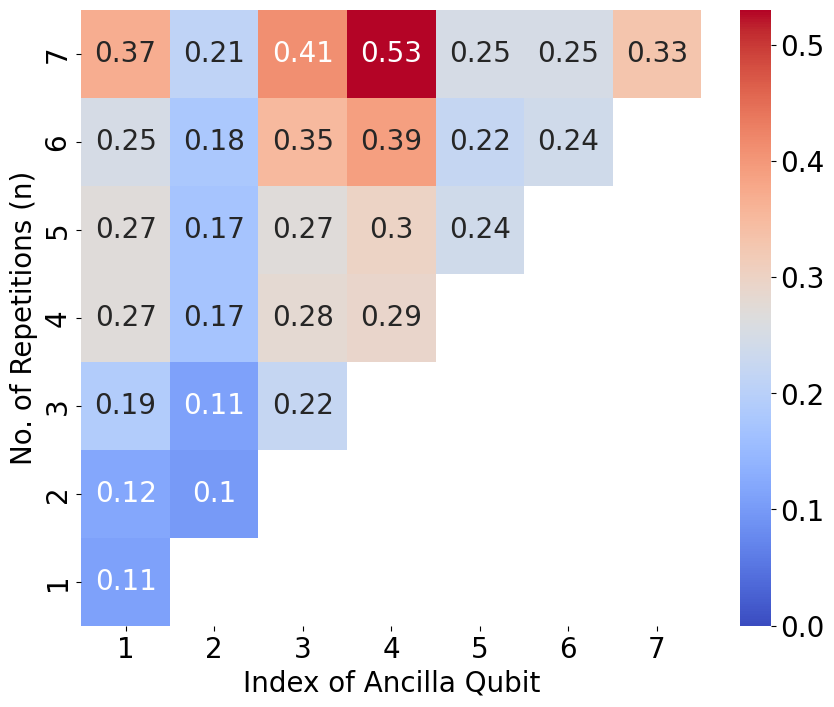}
        \caption{FRESH, optimization level 0}
        \label{subfig:ibm_fresh_opt0}
    \end{subfigure}
    \hfill
    \begin{subfigure}[t]{0.31\textwidth}
        \centering
        \includegraphics[width=\linewidth]{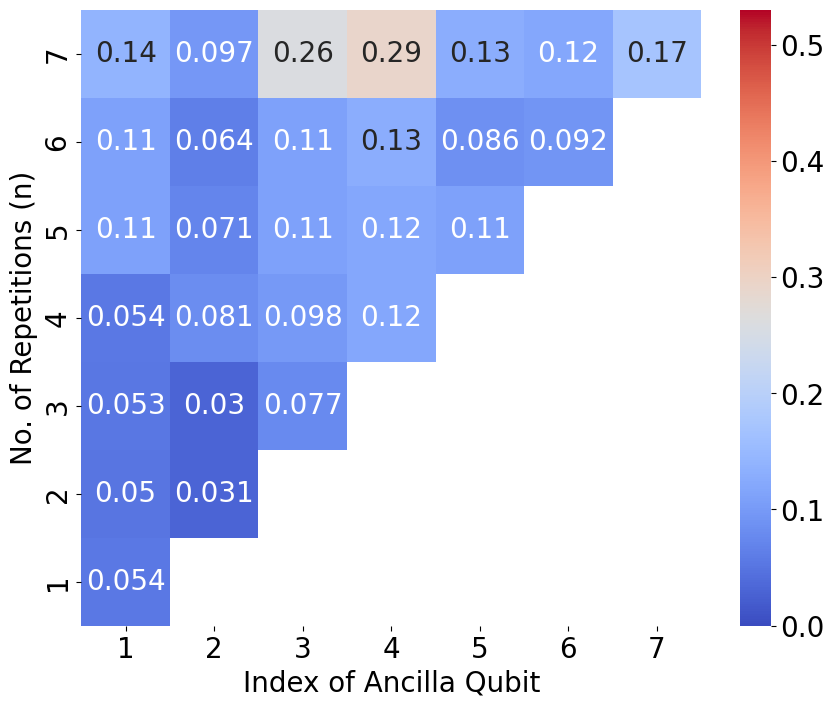}
        \caption{FRESH, optimization level 1}
        \label{subfig:ibm_fresh_opt1}
    \end{subfigure}
    \hfill
    \begin{subfigure}[t]{0.31\textwidth}
        \centering
        \includegraphics[width=\linewidth]{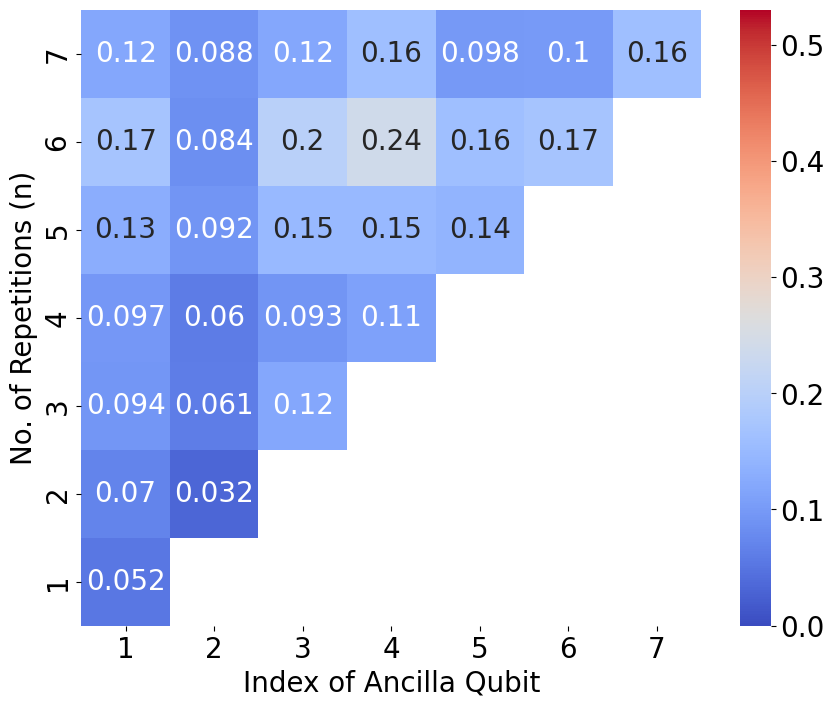}
        \caption{FRESH, optimization level 2}
        \label{subfig:ibm_fresh_opt2}
    \end{subfigure}

    \vspace{1em}

    \begin{subfigure}[t]{0.45\textwidth}
        \centering
        \includegraphics[width=\linewidth]{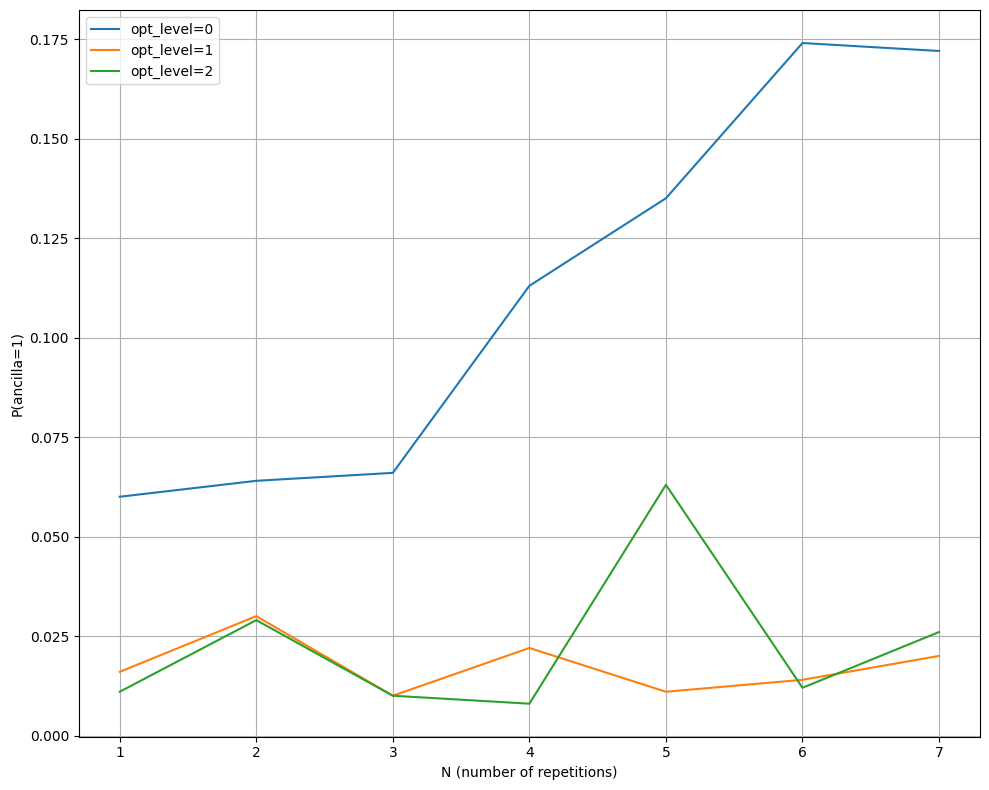}
        \caption{RECYCLE}
        \label{subfig:ibm_recycle}
    \end{subfigure}

    \caption{Experiments on IBM Torino.
    Probability that the ancilla measurement outcome is $1$ versus circuit depth.
    \textbf{(a)--(c)} FRESH strategy under transpiler optimization levels 0, 1, and 2.
    \textbf{(d)} RECYCLE strategy under the same optimization levels.}
    \label{fig:ibm_results}
\end{figure*}

For IBM Torino (Heron1 chip), we execute both FRESH and RECYCLE (Figure~\ref{fig:ibm_results}) and get the probability that the ancilla measurement outcome is $1$ as a function of circuit depth under IBM transpiler optimization levels 0, 1, and 2. All circuits were run with 1000 shots.

%% file: __main.bib
@mastersthesis{do2024automatic,
  title={Automatic Hardware-aware Optimization of Fault-tolerant Quantum Circuits},
  author={Do, Huyen},
  type={Bachelor's Thesis},
  school={Aalto University},
  url={https://urn.fi/URN:NBN:fi:aalto-202406184543},
  year={2024}
}

@article{Maslov_2017,
   title={Basic circuit compilation techniques for an ion-trap quantum machine},
   volume={19},
   ISSN={1367-2630},
   DOI={10.1088/1367-2630/aa5e47},
   number={2},
   journal={New Journal of Physics},
   publisher={IOP Publishing},
   author={Maslov, Dmitri},
   year={2017},
   month=feb, pages={023035} }

@book{Nielsen_Chuang_2010, 
place={Cambridge}, 
title={Quantum Computation and Quantum Information: 10th Anniversary Edition}, 
publisher={Cambridge University Press}, 
author={Nielsen, Michael A. and Chuang, Isaac L.}, 
year={2010}}

@misc{IonQNativeGates,
  author = {{IonQ}},
  title = {Getting Started with Native Gates},
  year = {2024},
  url = {https://ionq.com/docs/getting-started-with-native-gates},
  note = {Accessed: 2024-02-15}
}

@article{Hietala_2021,
   title={A verified optimizer for Quantum circuits},
   volume={5},
   ISSN={2475-1421},
   DOI={10.1145/3434318},
   number={POPL},
   journal={Proceedings of the ACM on Programming Languages},
   publisher={Association for Computing Machinery (ACM)},
   author={Hietala, Kesha and Rand, Robert and Hung, Shih-Han and Wu, Xiaodi and Hicks, Michael},
   year={2021},
   month=jan, pages={1–29} }

@INPROCEEDINGS{1219016,
  author={Miller, D.M. and Maslov, D. and Dueck, G.W.},
  booktitle={Proceedings 2003. Design Automation Conference (IEEE Cat. No.03CH37451)}, 
  title={A transformation based algorithm for reversible logic synthesis}, 
  year={2003},
  volume={},
  number={},
  pages={318-323},
  doi={10.1145/775832.775915}}

@misc{chatterjee2023qpandora,
      title={Q-Pandora Unboxed: Characterizing Noise Resilience of Quantum Error Correction Codes}, 
      author={Avimita Chatterjee and Subrata Das and Swaroop Ghosh},
      year={2023},
      eprint={2308.02769},
      archivePrefix={arXiv},
      primaryClass={quant-ph}
}

@article{eisert2025mind,
  title={Mind the gaps: The fraught road to quantum advantage},
  author={Eisert, Jens and Preskill, John},
  journal={arXiv preprint arXiv:2510.19928},
  year={2025}
}

@misc{preskill2025beyond,
  title={Beyond nisq: The megaquop machine},
  author={Preskill, John},
  journal={ACM Transactions on Quantum Computing},
  volume={6},
  number={3},
  pages={1--7},
  year={2025},
  publisher={ACM New York, NY}
}

@article{schoenberger2024shuttling,
  title={Shuttling for scalable trapped-ion quantum computers},
  author={Schoenberger, Daniel and Hillmich, Stefan and Brandl, Matthias and Wille, Robert},
  journal={IEEE Transactions on Computer-Aided Design of Integrated Circuits and Systems},
  volume={44},
  number={6},
  pages={2144--2155},
  year={2024},
  publisher={IEEE}
}

@article{kreppel2023quantum,
  title={Quantum circuit compiler for a shuttling-based trapped-ion quantum computer},
  author={Kreppel, Fabian and Melzer, Christian and Mill{\'a}n, Diego Olvera and Wagner, Janis and Hilder, Janine and Poschinger, Ulrich and Schmidt-Kaler, Ferdinand and Brinkmann, Andr{\'e}},
  journal={Quantum},
  volume={7},
  pages={1176},
  year={2023},
  publisher={Verein zur F{\"o}rderung des Open Access Publizierens in den Quantenwissenschaften}
}

@article{preskill1997faulttolerant,
  title={Fault-tolerant quantum computation},
  author={Preskill, John},
  journal={arXiv preprint quant-ph/9712048},
  year={1997}
}

@article{Fowler_2012,
   title={Surface codes: Towards practical large-scale quantum computation},
   volume={86},
   ISSN={1094-1622},
   DOI={10.1103/physreva.86.032324},
   number={3},
   journal={Physical Review A},
   publisher={American Physical Society (APS)},
   author={Fowler, Austin G. and Mariantoni, Matteo and Martinis, John M. and Cleland, Andrew N.},
   year={2012},
   month=sep }

@article{Trout_2018,
   title={Simulating the performance of a distance-3 surface code in a linear ion trap},
   volume={20},
   ISSN={1367-2630},
   DOI={10.1088/1367-2630/aab341},
   number={4},
   journal={New Journal of Physics},
   publisher={IOP Publishing},
   author={Trout, Colin J and Li, Muyuan and Gutiérrez, Mauricio and Wu, Yukai and Wang, Sheng-Tao and Duan, Luming and Brown, Kenneth R},
   year={2018},
   month=apr, pages={043038} }

@misc{quera,
    author = {{Quera.com}},
    title = {Trapped Ions},
    howpublished = {Online},
    year = {Accessed 2024},
    url = {https://www.quera.com/glossary/trapped-ions}
}

@article{erhard2021entangling,
  title={Entangling logical qubits with lattice surgery},
  author={Erhard, Alexander and Poulsen Nautrup, Hendrik and Meth, Markus and et al.},
  journal={Nature},
  volume={589},
  pages={220--224},
  year={2021},
  doi={10.1038/s41586-020-03079-6}
}

@Inbook{Kockum2019,
    author="Kockum, Anton Frisk
    and Nori, Franco",
    editor="Tafuri, Francesco",
    title="Quantum Bits with Josephson Junctions",
    bookTitle="Fundamentals and Frontiers of the Josephson Effect",
    year="2019",
    publisher="Springer International Publishing",
    address="Cham",
    pages="703--741",
    isbn="978-3-030-20726-7",
    doi="10.1007/978-3-030-20726-7_17",
}

@misc{wu2021mapping,
      title={Mapping Surface Code to Superconducting Quantum Processors}, 
      author={Anbang Wu and Gushu Li and Hezi Zhang and Gian Giacomo Guerreschi and Yufei Ding and Yuan Xie},
      year={2021},
      eprint={2111.13729},
      archivePrefix={arXiv},
      primaryClass={quant-ph}
}

@misc{ionq_backends,
  title = {IonQ Backends},
  howpublished = {\url{https://cloud.ionq.com/backends}},
  note = {Accessed: April 23, 2024}
}

@inproceedings{Leblond_2023, series={SC-W 2023},
   title={TISCC: A Surface Code Compiler and Resource Estimator for Trapped-Ion Processors},
   DOI={10.1145/3624062.3624214},
   booktitle={Proceedings of the SC ’23 Workshops of The International Conference on High Performance Computing, Network, Storage, and Analysis},
   publisher={ACM},
   author={Leblond, Tyler and Bennink, Ryan S. and Lietz, Justin G. and Seck, Christopher M.},
   year={2023},
   month=nov, collection={SC-W 2023} }

@INPROCEEDINGS {dummies2023,
author = {A. Chatterjee and K. Phalak and S. Ghosh},
booktitle = {2023 IEEE International Conference on Quantum Computing and Engineering (QCE)},
title = {Quantum Error Correction For Dummies},
year = {2023},
volume = {},
issn = {},
pages = {70-81},
doi = {10.1109/QCE57702.2023.00017},
publisher = {IEEE Computer Society},
}

@inproceedings{Wahl_2023,
   title={Zero Noise Extrapolation on Logical Qubits by Scaling the Error Correction Code Distance},
   DOI={10.1109/qce57702.2023.00103},
   booktitle={2023 IEEE International Conference on Quantum Computing and Engineering (QCE)},
   publisher={IEEE},
   author={Wahl, Misty A. and Mari, Andrea and Shammah, Nathan and Zeng, William J. and Ravi, Gokul Subramanian},
   year={2023},
   month=sep }

@book{gottesman1997stabilizer,
  title={Stabilizer codes and quantum error correction},
  author={Gottesman, Daniel},
  year={1997},
  publisher={California Institute of Technology}
}

@article{Derks_2025,
   title={Designing fault-tolerant circuits using detector error models},
   volume={9},
   ISSN={2521-327X},
   DOI={10.22331/q-2025-11-06-1905},
   journal={Quantum},
   publisher={Verein zur Forderung des Open Access Publizierens in den Quantenwissenschaften},
   author={Derks, Peter-Jan H.S. and Townsend-Teague, Alex and Burchards, Ansgar G. and Eisert, Jens},
   year={2025},
   month=nov, pages={1905} }

@article{zhang_2023,
  title = {Concatenation of the Gottesman-Kitaev-Preskill code with the XZZX surface code},
  author = {Zhang, Jiaxuan and Wu, Yu-Chun and Guo, Guo-Ping},
  journal = {Phys. Rev. A},
  volume = {107},
  issue = {6},
  pages = {062408},
  numpages = {16},
  year = {2023},
  month = {Jun},
  publisher = {American Physical Society},
  doi = {10.1103/PhysRevA.107.062408},
}

@article{ktb3-gcxr,
  title = {Logical error rates for the surface code under a mixed coherent and stochastic circuit-level noise model inspired by trapped ions},
  author = {LeBlond, Tyler and Groszkowski, Peter and Lietz, Justin G. and Seck, Christopher M. and Bennink, Ryan S.},
  journal = {Phys. Rev. Res.},
  volume = {7},
  issue = {4},
  pages = {043184},
  numpages = {18},
  year = {2025},
  month = {Nov},
  publisher = {American Physical Society},
  doi = {10.1103/ktb3-gcxr},
}

@article{gidney2021stim,
  doi = {10.22331/q-2021-07-06-497},
  url = {https://doi.org/10.22331/q-2021-07-06-497},
  title = {Stim: a fast stabilizer circuit simulator},
  author = {Gidney, Craig},
  journal = {{Quantum}},
  issn = {2521-327X},
  publisher = {{Verein zur F{\"{o}}rderung des Open Access Publizierens
                in den Quantenwissenschaften}},
  volume = 5,
  pages = 497,
  month = jul,
  year = 2021
}

@misc{IBMQuantumTorino,
  author       = {{IBM Quantum}},
  title        = {IBM Quantum Platform: ibm\_torino (Compute resources)},
  howpublished = {\url{https://quantum.cloud.ibm.com/?computer=ibm_torino}},
  note         = {Accessed: 2026-02-25}
}

@article{Ashkin1986SingleBeamOpticalTrap,
  author  = {Ashkin, A. and Dziedzic, J. M. and Bjorkholm, J. E. and Chu, Steven},
  title   = {Observation of a single-beam gradient force optical trap for dielectric particles},
  journal = {Optics Letters},
  volume  = {11},
  number  = {5},
  pages   = {288--290},
  year    = {1986},
  doi     = {10.1364/OL.11.000288}
}

@article{PhysRevLett.74.4091,
  title = {Quantum Computations with Cold Trapped Ions},
  author = {Cirac, J. I. and Zoller, P.},
  journal = {Phys. Rev. Lett.},
  volume = {74},
  issue = {20},
  pages = {4091--4094},
  numpages = {0},
  year = {1995},
  month = {May},
  publisher = {American Physical Society},
  doi = {10.1103/PhysRevLett.74.4091},
}

@article{Bluvstein2024LogicalQuantumProcessor,
author = {Bluvstein, Dolev and Evered, Simon J. and Geim, Aziza A. and et al.},
title = {Logical quantum processor based on reconfigurable atom arrays},
journal = {Nature},
volume = {626},
pages = {58--65},
year = {2024},
doi = {10.1038/s41586-023-06927-3}
}

@article{pinoDemonstrationTrappedionQuantum2021a,
  title = {Demonstration of the Trapped-Ion Quantum {{CCD}} Computer Architecture},
  author = {Pino, J. M. and Dreiling, J. M. and Figgatt, C. and Gaebler, J. P. and Moses, S. A. and Allman, M. S. and Baldwin, C. H. and {Foss-Feig}, M. and Hayes, D. and Mayer, K. and {Ryan-Anderson}, C. and Neyenhuis, B.},
  year = 2021,
  month = apr,
  journal = {Nature},
  volume = {592},
  number = {7853},
  pages = {209--213},
  publisher = {Nature Publishing Group},
  issn = {1476-4687},
  doi = {10.1038/s41586-021-03318-4},
  urldate = {2026-06-02},
  copyright = {2021 The Author(s), under exclusive licence to Springer Nature Limited part of Springer Nature},
  langid = {english}
}

@article{corcolesExploitingDynamicQuantum2021,
  title = {Exploiting {{Dynamic Quantum Circuits}} in a {{Quantum Algorithm}} with {{Superconducting Qubits}}},
  author = {C{\'o}rcoles, A. D.},
  year = 2021,
  journal = {Physical Review Letters},
  volume = {127},
  number = {10},
  doi = {10.1103/PhysRevLett.127.100501}
}

@article{proctorDetectingTrackingDrift2020,
  title = {Detecting and Tracking Drift in Quantum Information Processors},
  author = {Proctor, Timothy and Revelle, Melissa and Nielsen, Erik and Rudinger, Kenneth and Lobser, Daniel and Maunz, Peter and {Blume-Kohout}, Robin and Young, Kevin},
  year = 2020,
  month = oct,
  journal = {Nature Communications},
  volume = {11},
  number = {1},
  pages = {5396},
  issn = {2041-1723},
  doi = {10.1038/s41467-020-19074-4},
  urldate = {2026-06-03},
  langid = {english}
}

@article{Huo_2017,
   title={Learning time-dependent noise to reduce logical errors: real time error rate estimation in quantum error correction},
   volume={19},
   ISSN={1367-2630},
   url={http://dx.doi.org/10.1088/1367-2630/aa916e},
   DOI={10.1088/1367-2630/aa916e},
   number={12},
   journal={New Journal of Physics},
   publisher={IOP Publishing},
   author={Huo, Ming-Xia and Li, Ying},
   year={2017},
   month=Dec, pages={123032} }

@article{bruzewiczTrappedionQuantumComputing2019,
  title = {Trapped-Ion Quantum Computing: {{Progress}} and Challenges},
  shorttitle = {Trapped-Ion Quantum Computing},
  author = {Bruzewicz, Colin D. and Chiaverini, John and McConnell, Robert and Sage, Jeremy M.},
  year = 2019,
  month = may,
  journal = {Applied Physics Reviews},
  volume = {6},
  number = {2},
  pages = {021314},
  issn = {1931-9401},
  doi = {10.1063/1.5088164},
  urldate = {2026-06-03},
}

@article{zhengMinimizingReadoutinducedNoise2024,
  title = {Minimizing Readout-Induced Noise for Early Fault-Tolerant Quantum Computers},
  author = {Zheng, Yunzhe},
  year = 2024,
  journal = {Physical Review Research},
  volume = {6},
  number = {2},
  doi = {10.1103/PhysRevResearch.6.023129}
}
